\documentclass[prd,aps,showpacs,preprintnumbers,nofootinbib,eqsecnum, preprintnumbers]{revtex4}
\usepackage{graphicx}
\usepackage{epsf}
\usepackage{amsmath}
\usepackage{epstopdf}
\usepackage{bm}
\usepackage{color}
\usepackage{tabularx}
\usepackage{enumitem}
\usepackage{float}
\usepackage{array,booktabs}
\usepackage{footnote}
\usepackage{threeparttable}
\usepackage{graphicx}
\usepackage{hyperref}
\usepackage{amssymb,epsf}
\usepackage{latexsym}
\usepackage{epstopdf}
\usepackage{epsfig}

\begin{document}

\title{Critical Phenomena and Reentrant Phase Transition of \\
Asymptotically Reissner--Nordstr\"{o}m Black Holes}
\author{Mehrab Momennia$^{1,2}$\footnote{%
email address: momennia1988@gmail.com} and Seyed Hossein Hendi$^{1,2,3}$%
\footnote{%
email address: hendi@shirazu.ac.ir}} \affiliation{$^1$Department
of Physics, School of Science, Shiraz University, Shiraz
71454, Iran \\
$^2$Biruni Observatory, School of Science, Shiraz University, Shiraz 71454,
Iran \\
$^{3}$Canadian Quantum Research Center 204-3002 32 Ave Vernon, BC V1T 2L7
Canada }

\begin{abstract}
By considering a small correction to the Maxwell field, we show that the
resultant black hole solutions (also known as the asymptotically
Reissner--Nordstr\"{o}m black holes) undergo the reentrant phase transition
and can have a novel phase behavior. We also show that such a small
nonlinear correction of the Reissner--Nordstr\"{o}m black holes has high
effects on the phase structure of the solutions. It leads to a new
classification in the canonical ensemble of extended phase space providing
the values of the nonlinearity parameter $\alpha$ being $\alpha \lesseqqgtr
4q^{2}/7$. We shall study these three classes and investigate deviations
from those of the standard Reissner--Nordstr\"{o}m solutions. Interestingly,
we find that there is the reentrant phase transition for $\alpha <4q^{2}/7$,
and for the case of $\alpha =4q^{2}/7$ there is no phase transition below
(at) the critical point. For the last case, one finds that small and large
black holes are thermodynamically distinguishable for temperatures and
pressures higher than the critical ones.
\end{abstract}

\maketitle


\section{Introduction}

It is well known that a black hole can be investigated as an ordinary
thermodynamic system \cite{Davies,Davies1989,Wald} with typical entropy \cite%
{Bekenstein} and temperature \cite{Hawking} such that in most cases usually
obeys the first law of thermodynamics \cite{Bardeen}. It was also shown that
these highly dense compact objects treat as usual thermodynamic systems
enjoying the phase transition phenomenon \cite{HP}. More interestingly, we
can see the van der Waals-like (vdW-like) phase transition including charged
black hole systems by considering the correspondence $\left( Q,\Phi \right)
\leftrightarrow \left( P,V\right) $\ between conserved quantities and
thermodynamic variables \cite{Chamblin,ChamblinEmparan}. Recently, in the
context of black hole thermodynamics, a possible connection between the
cosmological constant as a thermodynamical pressure is proposed \cite%
{Caldarelli,Kastor} which has attracted much attention. This relation is
defined as follows
\begin{equation}
P=-\frac{\Lambda }{8\pi },  \label{P2}
\end{equation}%
in which the thermodynamical volume $V$\ is the conjugate quantity to
pressure\ as
\begin{equation}
V=\left( \frac{\partial M}{\partial P}\right) _{rep},
\end{equation}
where "$rep$" refers to "residual~extensive parameters". Indeed, the primary
motivation of considering $\Lambda $ as a thermodynamical pressure comes
from the fact that several physical constants, such as Yukawa coupling,
gauge coupling constants, and Newton's constant are not fixed values in some
fundamental theories. In addition, in Tolman--Oppenheimer--Volkoff equation,
$\Lambda $ is added to pressure that shows the cosmological constant can
play the role of the thermodynamical pressure. Besides, $\Lambda $ is a slow
variation parameter and has the dimension $(length)^{-2}$ which is the
dimension of the pressure. Usually, a vdW-like small-large black hole
(SBH-LBH) phase transition can be observed in thermodynamical systems
including black holes whenever $\Lambda $ behaves as a thermodynamical
pressure. This type of phase transition has been studied extensively in the
background spacetime of various black hole solutions (for instance, see an
incomplete list \cite%
{KubiznakMann,Banerjee,Wei,Mo,Zou,Xu,HendiFaizal,Mandal,Miao,RainbowYM,StetskoPRD,MassiveYM,Estrada,Stetsko2021}
and references therein written during recent years).

The reentrant phase transition (RPT) phenomenon can be observed in
an ordinary thermodynamical system when a monotonic change of any
thermodynamical variable provides more than one phase transition
such that the final phase is macroscopically similar to the
initial phase. There is a special range in temperature in the
asymptotically Reissner--Nordstr\"{o}m (ARN) black holes so that
these solutions enjoy a large-small-large phase transition by a
monotonic change in the pressure. This interesting phase behavior
has been observed in ordinary thermodynamical systems, such as
nicotine-water mixture \cite{Hudson}, liquid crystals, binary
gases, multicomponent fluids, and other different typical
thermodynamic systems \cite{Narayanan}. In the context of black
hole thermodynamics, the RPT is reported for Born-Infeld solutions
\cite{BIadS,AminRPT}, rotating black holes \cite{rotatingadS},
asymptotically dS black holes \cite{deSitter},
hairy black holes \cite{hairy}, black hole solutions in massive gravity \cite%
{dRGTmassive}, and Born--Infeld-dilaton black holes \cite{reentrant}.

In this paper, we study the thermodynamics of ARN black holes, investigate
the RPT in the extended phase space, and find deviations from those of the
standard Reissner--Nordstr\"{o}m (RN) solutions. We also discuss novel
phenomenon of our black hole case study and compare it with the standard RN
black holes.


\section{review of solutions and thermodynamics \label{FE}}

In this section, we are going to briefly mention the solutions and
thermodynamics of black holes in the presence of quadratic nonlinear
electrodynamics. Before proceeding, it is worthwhile to give some
motivations.\newline
Nonlinear field theories are of interest in various classes of mathematical
physics since most physical systems are basically nonlinear in the nature.
The nonlinear electrodynamic (NED) fields are much richer than the Maxwell
theory and in special cases they reduce to the linear Maxwell field.
Different constraints of the Maxwell field, like the radiation propagation
inside specific materials \cite{NLmaterial,Lorenci,Novello,Bittencourt} and
description of the self-interaction of virtual electron-positron pairs \cite%
{H-E,Yajima,Schwinger}, motivate one to consider NED theories as effective
fields \cite{DelphenichQED,Delphenich}. Moreover, a well-known outstanding
problem is that most gravitational theories predict a singularity in the
center of black holes. It was shown that by employing the NED fields, the
big bang and black hole singularities can be removed \cite%
{bigbang,Ayon,Klippert,Dymnikova,Corda,Cuesta}. Besides, the NEDs have
important effects on the structure of the superstrongly magnetized compact
objects, such as pulsars and strange stars \cite{Birula,Salim,CuestaSalim}.

The Lagrangian of Born-Infeld-type NED theories \cite%
{Born,Infeld,Soleng,Hendi}, which each one was constructed based on various
motivations,\ tends to the following form for weak nonlinearity \cite%
{Topologicalinstability}%
\begin{equation}
L(F)=-F+\alpha F^{2}+O\left( \alpha ^{2}\right) ,  \label{Lagrangian}
\end{equation}%
where $F=F_{\mu \nu }F^{\mu \nu }$ is the Maxwell invariant, $F_{\mu \nu
}=\partial _{\mu }A_{\nu }-\partial _{\nu }A_{\mu }$ is the electromagnetic
field tensor, and $A_{\mu }$ is the gauge potential. In this equation, $%
\alpha $ denotes nonlinearity parameter that is a small quantity and
proportional to the inverse value of nonlinearity parameter in
Born-Infeld-type NED fields. Indeed, although different models of NEDs have
been constructed with various primitive aims, they contain physical and
experimental importance just for the weak nonlinearity since the Maxwell
field in various branches leads to near accurate or acceptable results.
Thus, in transition from the Maxwell field to NEDs, considering the weak
nonlinearity effects seems to be reasonable and a logical decision. In other
words, we expect to obtain precise physical results with experimental
agreements whenever the nonlinearity is considered as a correction to the
Maxwell theory. In this context, regardless of constant parameters which are
contracted in $\alpha $, most NED Lagrangians reduce to Eq. (\ref{Lagrangian}%
) for weak nonlinearity and we shall consider this Lagrangian as an
effective matter source coupled to gravity.

The mentioned motivations have led to publish some interesting and
reasonable works by employing Eq. (\ref{Lagrangian}) as an effective
Lagrangian of electrodynamics \cite%
{H-E,Yajima,Schwinger,Stehle,DelphenichQED,Delphenich,Kats,Nie,Anninos,BIString,Fradkin,Matsaev,Pope,Tseytlin,Gross}%
. Heisenberg and Euler demonstrated that quantum corrections lead to
nonlinear properties of vacuum \cite%
{H-E,Yajima,Schwinger,Stehle,DelphenichQED,Delphenich}. Besides, it was
shown that a quartic correction of the Maxwell invariant appears in the
low-energy limit of heterotic string theory \cite%
{Kats,Nie,Anninos,BIString,Fradkin,Matsaev,Pope,Tseytlin,Gross}. Therefore,
considering a correction term to the Maxwell field and investigating Eq. (%
\ref{Lagrangian}) as an effective and suitable Lagrangian of electrodynamics
instead of the Maxwell and other NED fields is a reasonable and logical
decision.

According to the mentioned motivations, we consider the topological black
holes in $(n+1)$-dimensional spacetime with perturbative nonlinear
electrodynamics \cite{Topologicalinstability}. The $(n+1)$-dimensional line
element reads
\begin{equation}
ds^{2}=-f(r)dt^{2}+\frac{dr^{2}}{f(r)}+r^{2}d\Omega _{n-1}^{2},
\label{metric}
\end{equation}
where $f(r)$ is the metric function and $d\Omega _{n-1}^{2}$ represents the
line element of $\left( n-1\right) $-dimensional hypersurface with constant
curvature $\left( n-1\right) \left( n-2\right) k$ and volume $\omega_{n-1}$
with the following explicit form%
\begin{equation}
d\Omega _{n-1}^{2}=\left\{
\begin{array}{cc}
d\theta _{1}^{2}+\sum\limits_{i=2}^{n-1}\prod\limits_{j=1}^{i-1}\sin
^{2}\theta _{j}d\theta _{i}^{2} & k=1 \\
d\theta _{1}^{2}+\sinh ^{2}\theta _{1}\left( d\theta
_{2}^{2}+\sum\limits_{i=3}^{n-1}\prod\limits_{j=2}^{i-1}\sin ^{2}\theta
_{j}d\theta _{i}^{2}\right) & k=-1 \\
\sum\limits_{i=1}^{n-1}d\phi _{i}^{2} & k=0%
\end{array}%
\right. ,
\end{equation}

The metric function of these black holes can be obtained as \cite%
{Topologicalinstability}%
\begin{equation}
f(r)=k-\frac{m}{r^{n-2}}-\frac{2\Lambda r^{2}}{n\left( n-1\right) }+\frac{%
2q^{2}}{\left( n-1\right) \left( n-2\right) r^{2n-4}}-\frac{4q^{4}}{\left[
2\left( n-2\right) \left( n+2\right) +\left( n-3\right) \left( n-4\right) %
\right] r^{4n-6}}\alpha +\mathcal{O}\left( \alpha ^{2}\right) ,  \label{MF}
\end{equation}%
in which $m$ and $q$\ are two integration constants which are related to the
total mass and total electric charge of the black hole, and the last term
indicates the effect of nonlinearity.

The Hawking temperature of these black holes can obtained~by using the
definition of the surface gravity on the outermost horizon, $r_{+}$,%
\begin{equation}
T=\frac{1}{2\pi \left( n-1\right) }\left( \frac{\left( n-1\right) \left(
n-2\right) k}{2r_{+}}-\Lambda r_{+}-\frac{q^{2}}{r_{+}^{2n-3}}+\frac{2q^{4}}{%
r_{+}^{4n-5}}\alpha \right) +\mathcal{O}\left( \alpha ^{2}\right) .
\label{T}
\end{equation}

Moreover, as we are working in Einstein gravity, the entropy of the black
holes can be calculated via the quarter of the event horizon area%
\begin{equation}
S=\frac{r_{+}^{n-1}}{4},  \label{S}
\end{equation}%
which shows the entropy per unit volume $\omega_{n-1}$. The electric
potential $\Phi $, measured at infinity as a reference with respect to the
event horizon is given by%
\begin{equation}
\Phi =\frac{q}{\left( n-2\right) r_{+}^{n-2}}-\frac{4q^{3}}{\left(
3n-4\right) r_{+}^{3n-4}}\alpha +\mathcal{O}\left( \alpha ^{2}\right) .
\end{equation}

Besides, the total electric charge per unit volume $\omega_{n-1}$, can be
obtained by considering the flux of the electric field at infinity as%
\begin{equation}
Q=\frac{q}{4\pi }.
\end{equation}

At the final stage of calculating the conserved and thermodynamic
quantities, one can get the total mass of obtained black holes by using the
behavior of the metric at large $r$. Therefore, the total mass per unit
volume $\omega_{n-1}$ is given by%
\begin{equation}
M=\frac{\left( n-1\right) m}{16\pi }.  \label{M}
\end{equation}

Considering the entropy and electric charge as a complete set of extensive
parameters, one can show that these conserved and thermodynamic quantities
satisfy the first law of thermodynamics \cite{Topologicalinstability}%
\begin{equation}
dM=TdS+\Phi dQ.  \label{FL}
\end{equation}

It is worthwhile to mention that all the equations (\ref{MF})-(\ref{FL}) are
representing the background geometry and thermodynamics of the higher
dimensional ARN black holes and they reduce to the higher dimensional
standard RN solutions in the special limit $\alpha =0$.

\section{REENTRANT PHASE TRANSITION \label{PV}}

It is proved that the Schwarzschild (AdS) black holes have no
vdW-like phase transition \cite{Zhang}, while this phenomenon has
been observed in the RN black holes \cite{KubiznakMann}. Thus, the
electric charge of black holes usually plays a key role in
observing such a phase transition and it would be interesting to
include the effects of black hole's charge in the thermodynamic
calculations. In this paper, we show how a small correction to the
Maxwell field highly affects the phase transition structure of the
RN solutions and extends the thermodynamical phase space into
three new different regions. The mentioned correction is motivated
by nonlinear properties of vacuum generated from quantum
corrections, appearing a quartic correction of the Maxwell
invariant in the low-energy limit of heterotic string theory, and
physical and experimental importance of adding a weak nonlinearity
to the Maxwell field.

In what follows, we concentrate our attention on the spherical symmetric
black holes with negative cosmological constant in $4$-dimensional
spacetime. Calculations show that the RPT can also occur for higher
dimensional solutions, but this is not the case for positive cosmological
constant and/or flat or hyperbolic solutions. The negative cosmological
constant in the extended phase space plays the role of a positive
thermodynamical pressure as follows \cite{Caldarelli,Kastor}%
\begin{equation}
P=-\frac{\Lambda }{8\pi }.  \label{P1}
\end{equation}

In this scenario, the total mass (\ref{M})\ behaves as the enthalpy of
system, and the Smarr formula and first law of thermodynamics are modified as%
\begin{equation}
M=2TS+\Phi Q-2VP+2\mathcal{A}\alpha ;\ \ \ \ \ \mathcal{A}=\left( \frac{%
\partial M}{\partial \alpha }\right) _{S,Q,P},
\end{equation}%
\begin{equation}
dM=TdS+\Phi dQ+VdP+\mathcal{A}d\alpha ,
\end{equation}%
where $\mathcal{A}$\ is a new thermodynamical variable conjugate to $\alpha $
and as mentioned before, $V$\ is the thermodynamical volume conjugate to $P$
as follows
\begin{equation}
V=\left( \frac{\partial M}{\partial P}\right) _{S,Q,\alpha }=\frac{1}{3}%
r_{+}^{3}.
\end{equation}

Here, we study the thermodynamics of $4$-dimensional black holes in the
canonical ensemble (fixed $Q$ and $\alpha $) of extended phase space. So, by
using the temperature (\ref{T}) for $n=3$%
\begin{equation}
T=-\frac{\Lambda r_{+}}{4\pi }+\frac{1}{4\pi r_{+}}-\frac{q^{2}}{4\pi
r_{+}^{3}}+\frac{q^{4}\alpha }{2\pi r_{+}^{7}},  \label{T3}
\end{equation}%
and the relation between the cosmological constant and pressure (\ref{P1}),
it is straightforward to show that the equation of state, $P=P\left(
r_{+},T\right) $,\ is given by%
\begin{equation}
P=\frac{T}{2r_{+}}-\frac{1}{8\pi r_{+}^{2}}+\frac{q^{2}}{8\pi r_{+}^{4}}-%
\frac{q^{4}\alpha }{4\pi r_{+}^{8}}.  \label{P}
\end{equation}%
\textbf{\ }
\begin{figure}[tbp]
$%
\begin{array}{ccc}
\epsfxsize=7.5cm \epsffile{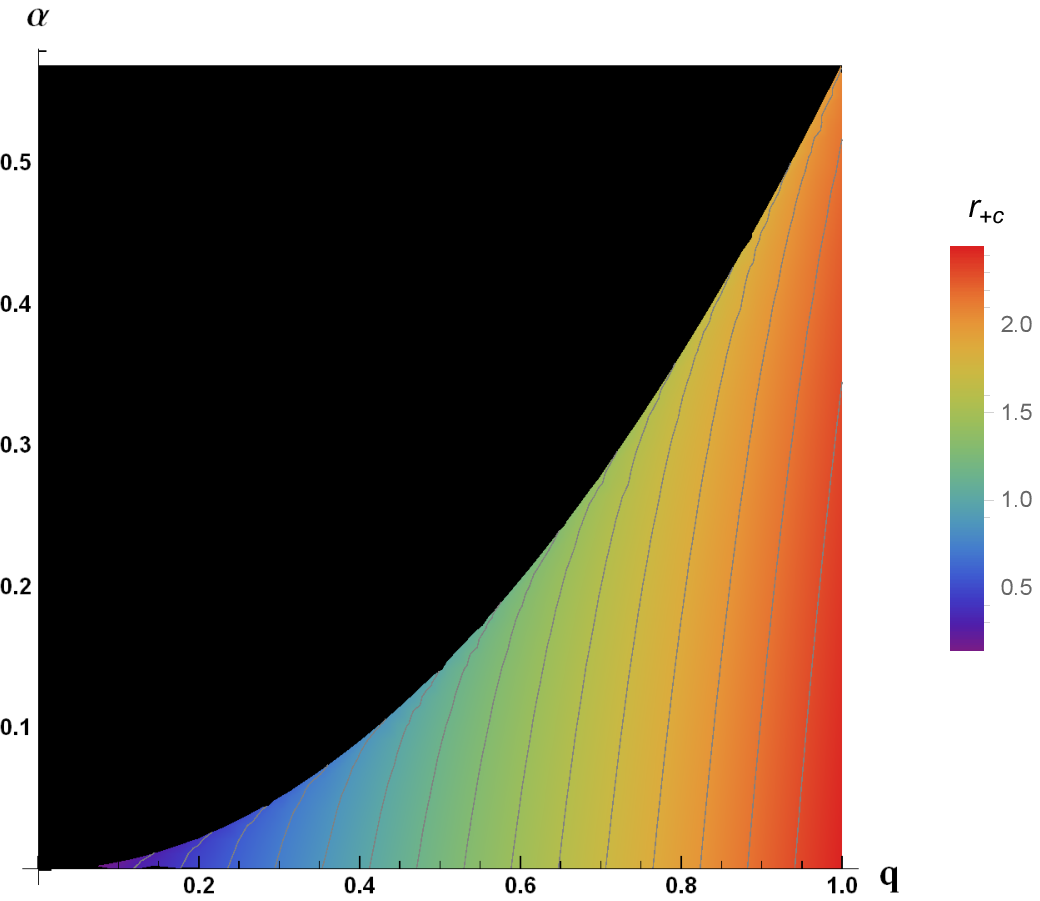} & \epsfxsize=7.5cm \epsffile{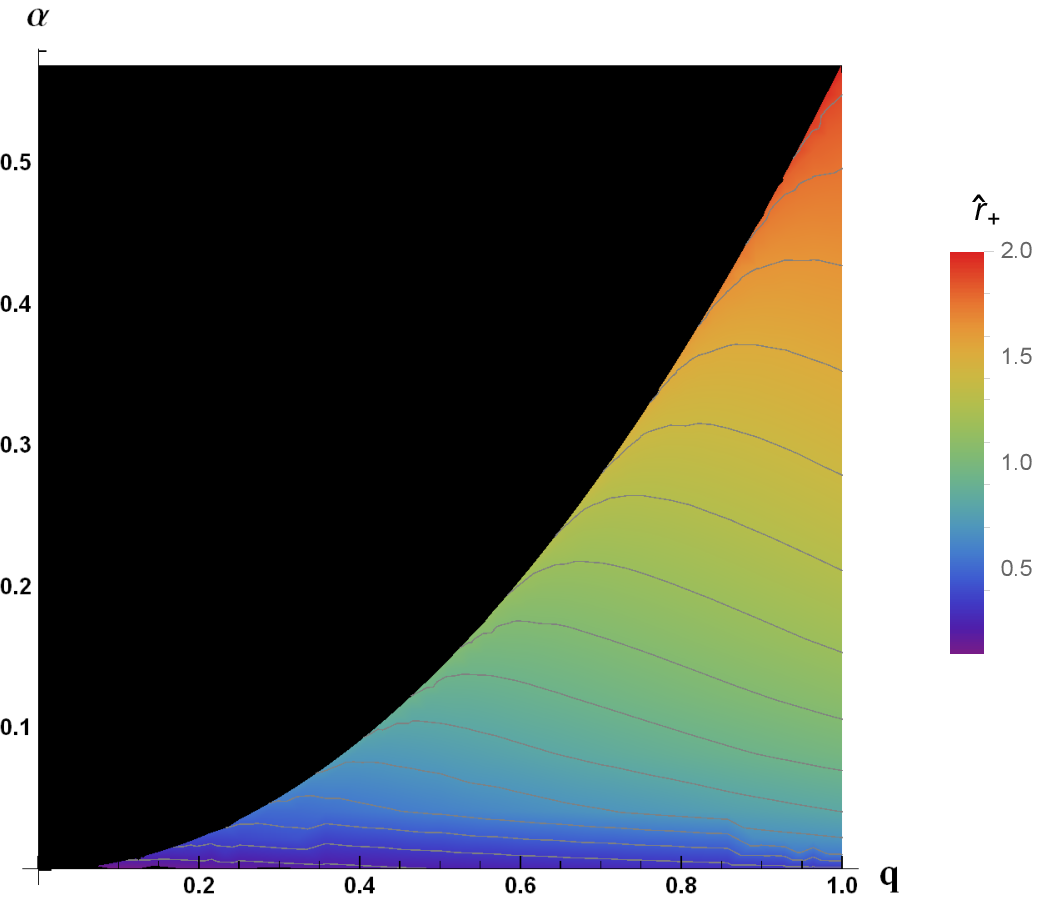} &
\end{array}
$%
\caption{$r_{+c}$ and $\hat{r}_{+}$ in $q-\protect\alpha$ plane. The black
region on the left indicates imaginary ${r}_{+}$ and $\hat{r}_{+}$ whereas
the colorful area represents real ${r}_{+}$ and $\hat{r}_{+}$. At the border
between black and colorful areas, ${r}_{+}$ and $\hat{r}_{+}$ are equal to $%
2q$.}
\label{rhat}
\end{figure}
\begin{figure}[tbp]
$%
\begin{array}{ccc}
\epsfxsize=7.5cm \epsffile{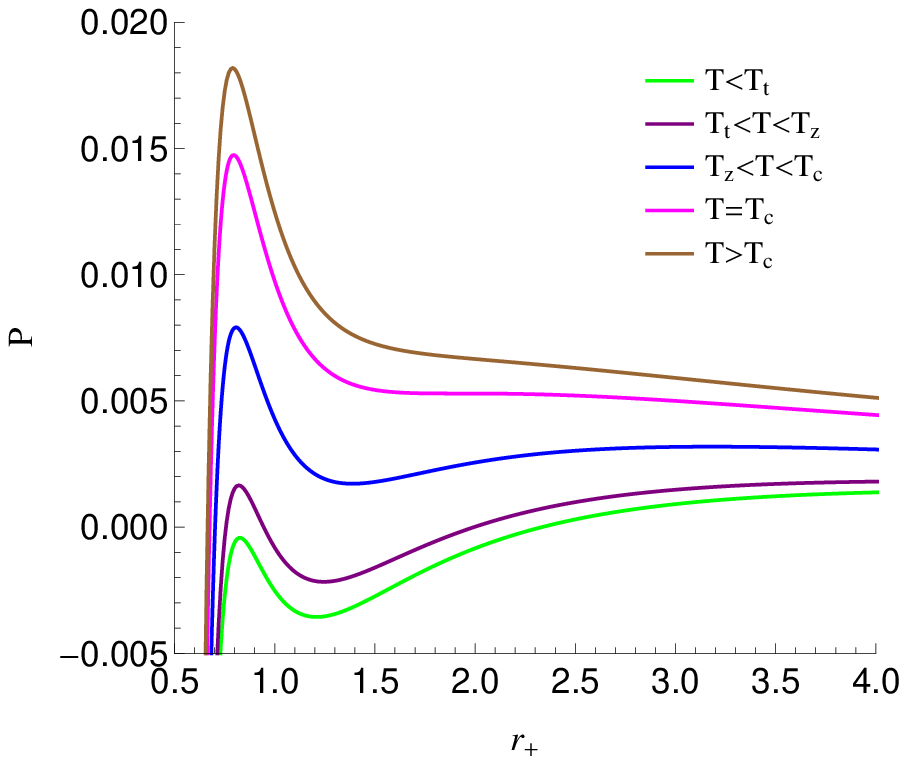} & \epsfxsize=7.5cm \epsffile{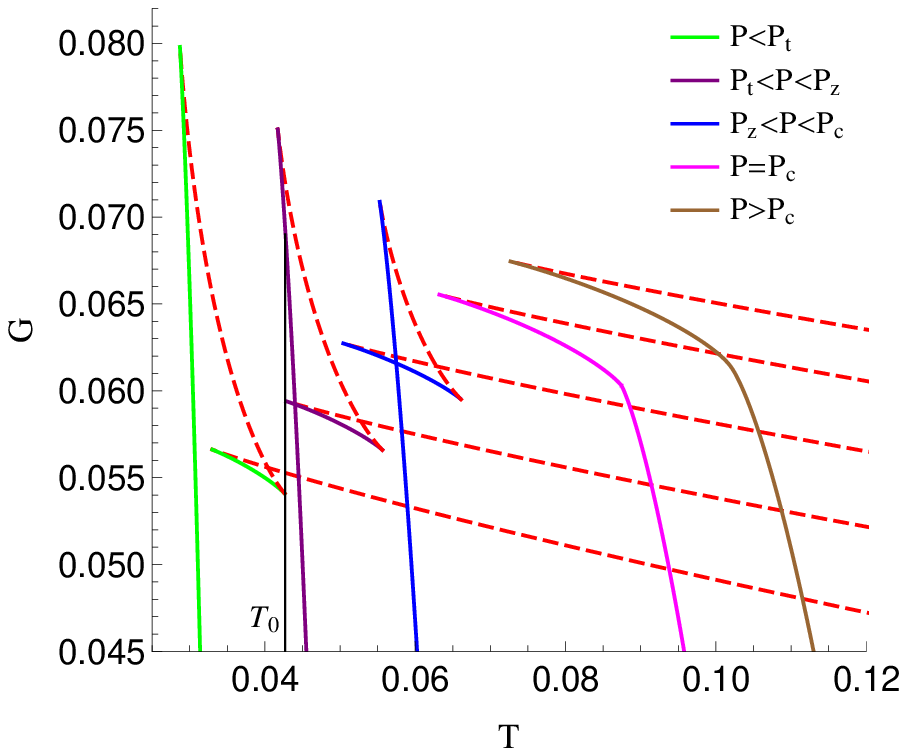} &
\end{array}
$%
\caption{$P-r_{+}$ and $G-T$ diagrams for different regions of
temperatures and pressures. In the right panel, the solid lines
refer to $C_{P}>0$ while the dashed red lines correspond to
$C_{P}<0$. $C_{P}$ diverges at the joins of dashed and solid
lines. Besides, $G-T$ curves are shifted for clarity. The vertical
black line at $T_{0}$, $T_{t}<T_{0}<T_{z}$, shows a discontinuity
in the Gibbs free energy and indicates a zeroth-order SBH-IBH
phase transition.} \label{PVfig}
\end{figure}
\begin{figure}[tbp]
$%
\begin{array}{ccc}
\epsfxsize=7.5cm \epsffile{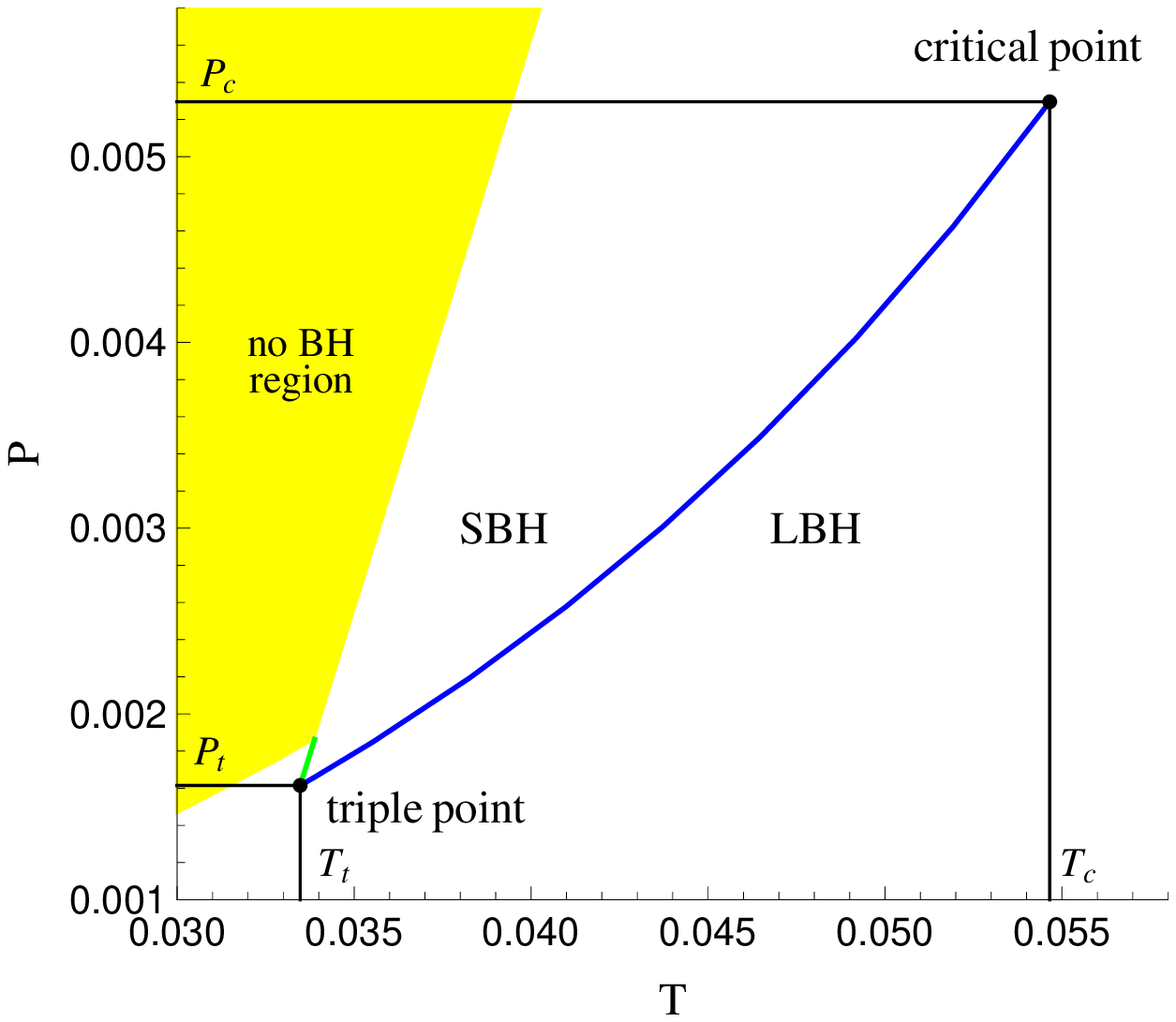} & \epsfxsize=7.5cm \epsffile{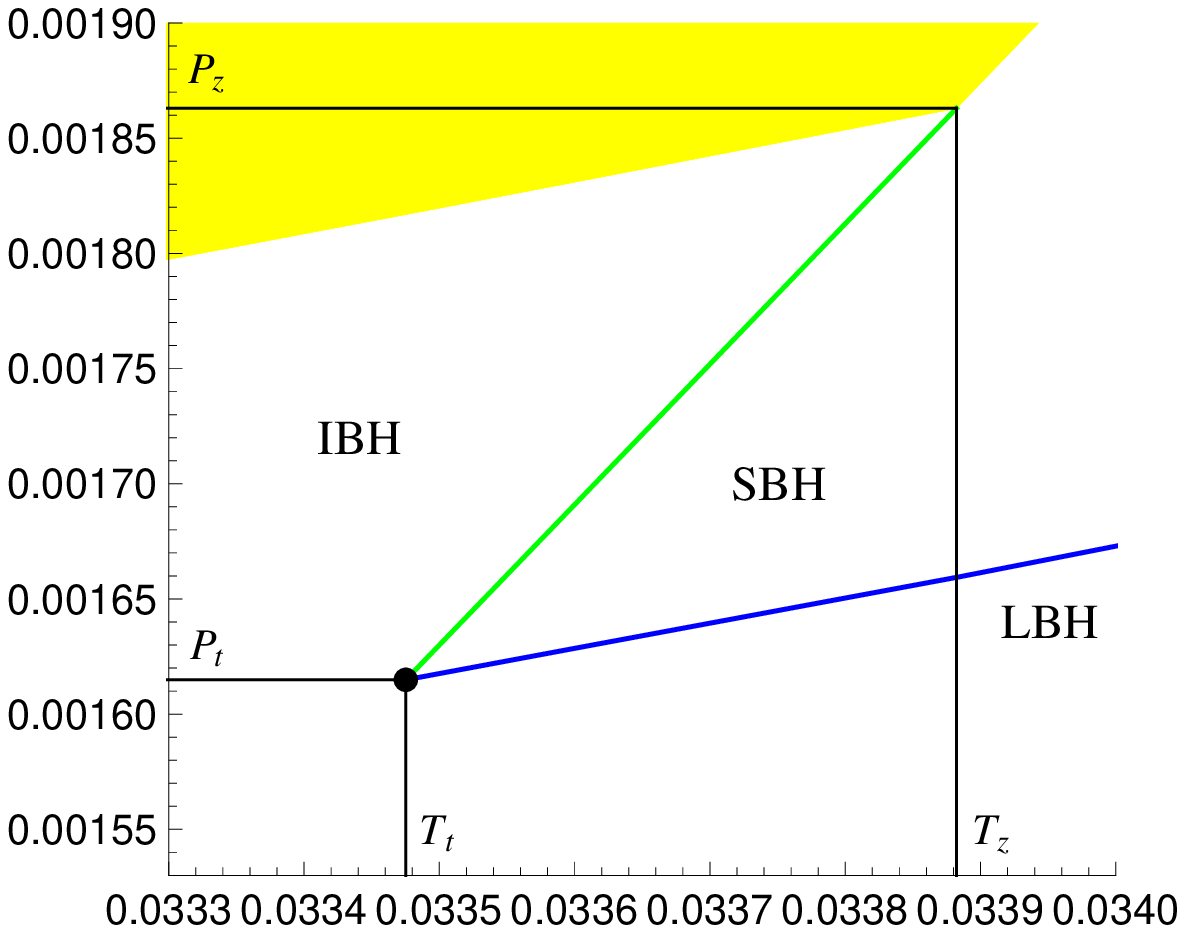} &
\end{array}
$%
\caption{$P-T$ diagram. The yellow region illustrates the no black
hole area and the right panel represents a close-up of the RPT
area. The blue curve shows the coexistence line of SBHs and LBHs
whereas the green curve refers to the coexistence line of IBHs and
small ones. On crossing the blue (green) line, the system goes
under a first (zeroth) -order phase transition between SBHs and
LBHs (IBHs and SBHs).} \label{PTfig}
\end{figure}

The thermodynamic behavior of the system and its global stability are
governed by the free energy, and thus, we obtain the Gibbs free energy as
well. We can determine the Gibbs free energy per unit volume $\omega _{2}$
in the extended phase space by employing the following relation%
\begin{equation}
G=M-TS=\frac{r_{+}}{16\pi }-\frac{r_{+}^{3}P}{6}+\frac{3q^{2}}{16\pi r_{+}}-%
\frac{7q^{4}\alpha }{40\pi r_{+}^{5}}.  \label{G}
\end{equation}

On the other hand, the heat capacity help us to find the local thermal
stability, and thus, we calculate it in extended phase space at constant
pressure as%
\begin{equation}
C_{P}=T\left( \frac{\partial S}{\partial T}\right) _{P}=\frac{%
r_{+}^{2}\left( 8\pi Pr_{+}^{8}+r_{+}^{6}-q^{2}r_{+}^{4}+2q^{4}\alpha
\right) }{2\left( 8\pi Pr_{+}^{8}-r_{+}^{6}+3q^{2}r_{+}^{4}-14q^{4}\alpha
\right) }.  \label{Cp}
\end{equation}

Here, since we are working in the canonical ensemble, $C_{P}$ is
the heat capacity at constant $P$, $Q$, and $\alpha $. The
negativity of $C_{P}$\ indicates unstable solutions while its
positivity refers to local stability (or at least metastability).

In order to study the phase transition of black holes, one can use the
definition of inflection point at the critical point of isothermal $P-V$ (or
equivalently $P-r_{+}$) diagram
\begin{equation}
\left. \frac{\partial P(r_{+},T)}{\partial r_{+}}\right\vert _{T}=\left.
\frac{\partial ^{2}P(r_{+},T)}{\partial r_{+}^{2}}\right\vert _{T}=0,
\label{IF}
\end{equation}%
which can be used to obtain the critical horizon radius $r_{+c}$ and
critical temperature $T_{c}$. One can easily show that this equation leads
to the following equation for the critical horizon radius%
\begin{equation}
r_{+}^{6}-6q^{2}r_{+}^{4}+56q^{4}\alpha =0,  \label{req}
\end{equation}%
with at most two real positive solutions as follows%
\begin{equation}
r_{+c}=\sqrt{2q^{2}\left( 1+\frac{q^{2}}{\mathcal{X}}\right) +2\mathcal{X}},
\label{rc}
\end{equation}%
\begin{equation}
\hat{r}_{+c}=\sqrt{2q^{2}\left( 1+\frac{i\left( i+\sqrt{3}\right) q^{2}}{2%
\mathcal{X}}\right) +i\left( i-\sqrt{3}\right) \mathcal{X}},  \label{rbar}
\end{equation}%
where%
\begin{equation}
\mathcal{X}=\left( q^{6}-\frac{7}{2}q^{4}\alpha +\frac{1}{2}\sqrt{7\alpha
q^{8}\left( 7\alpha -4q^{2}\right) }\right) ^{1/3}.
\end{equation}

From now, the thermodynamic behavior of these black holes depends on the
values of (\ref{rc})\ and (\ref{rbar}) which are illustrated in Fig. \ref%
{rhat}; (a) when $r_{+c}$\ and $\hat{r}_{+c}$\ are imaginary/complex, there
is neither vdW-like phase transition nor RPT (black region of Fig. \ref{rhat}%
). In addition, the behavior is not like the ideal gas and we shall discuss
this region later. (b) in the colorful area of Fig. \ref{rhat} that both $%
r_{+c}$ and $\hat{r}_{+c}$\ are real, the RPT is observed which we
investigate it in this section. (c) at the border of these black\ and
colorful areas, $r_{+c}$ is equal to $\hat{r}_{+c}$\ and is determined by $%
\alpha =4q^{2}/7$. In this case, there is a critical point such that no
phase transition occurs below and at this point. Besides, the SBHs and LBHs
are thermodynamically distinguishable above this critical point and we will
study this border in the next section. (d) for some higher values of $q$\
and $\alpha $, $\hat{r}_{+c}$\ is imaginary/complex while $r_{+c}$ is real.
This leads to the standard vdW-like (first-order SBH-LBH) phase transition
which is investigated extensively before (for instance, see \cite%
{KubiznakMann}\ for the standard RN black holes and \cite{vdW} for our black
hole case study) and we do not consider it in this paper. The other option,
means real $\hat{r}_{+c}$\ and imaginary $r_{+c}$,\ is not accessible for
the system.

One may note that in the absence of the nonlinearity (the case of RN black
hole), $r_{+c}$ reduces to $\sqrt{6}q$\ and $\hat{r}_{+c}$\ vanishes, as it
should be. Therefore, the RN black hole can only undergo the vdW-like phase
transition for nonzero values of the electric charge. This fact uncovers the
significant role of the nonlinearity parameter $\alpha $ on the phase
transition structure of these black holes. However, there is a constraint on
choosing the values of $q$\ and $\alpha $. Considering the last
(correction)\ term of Eqs. (\ref{P})\ and (\ref{G}), we should choose some
values of $q$\ and $\alpha $\ so that this term be ignorable compared with
the third term, and therefore, can be considered as a perturbation. Hence,
as the simplest option, we can consider some values of $q$ and $\alpha $ so
that $\alpha q^{2}<<r_{+}^{4}/2$. However, we can ignore this restriction
since one can consider the last term as a nonlinear term rather than just a
perturbation (correction) term.

As a typical example and without loss of generality, we consider $q=0.8$\
and $\alpha =0.1$\ that is a point in the colorful area of Fig. \ref{rhat},
and thus, we expect to see the RPT. Now, we can obtain the critical
temperature and pressure as follows%
\begin{equation}
T_{c}=\frac{1}{2\pi r_{+c}}-\frac{q^{2}}{\pi r_{+c}^{3}}+\frac{4q^{4}\alpha
}{\pi r_{+c}^{7}},  \label{tc}
\end{equation}%
\begin{equation}
P_{c}=\frac{T_{c}}{2r_{+c}}-\frac{1}{8\pi r_{+c}^{2}}+\frac{q^{2}}{8\pi
r_{+c}^{4}}-\frac{q^{4}\alpha }{4\pi r_{+c}^{8}}.  \label{pc}
\end{equation}

For the fixed $q=0.8$\ and $\alpha =0.1$, the general behavior of the ARN
black holes is shown in Fig. \ref{PVfig}. This figure is plotted for various
areas of temperature (or equivalently pressure) in $P-r_{+}$\ (or
equivalently $G-T$) diagram. The dashed red lines show the negative heat
capacity (\ref{Cp}) and refer to unstable black holes while the solid lines
stand for the positive heat capacity representing the stable (or metastable)
black holes (for more discussion regarding the relation between Gibbs free
energy and heat capacity see \cite{MassiveYM}). Considering Fig. \ref{PVfig}%
, we see that a critical point located at $P=P_{c}$ in $G-T$ diagram (at $%
T=T_{c}$ in $P-r_{+}$ diagram with an inflection point) and demonstrates a
second-order phase transition from SBHs to large ones. In $G-T$ diagram, the
curve looks like the Hawking-Page phase transition for $P>P_{c}$ \cite{HP}.
For $P_{t}<P<P_{c}$\ and $T_{t}<T<T_{c}$, there is an area that black holes
undergo the standard first-order SBH-LBH phase transition. Besides, there
are three different phases including intermediate black holes (IBHs), SBHs,
and LBHs for $P\in \left( P_{t},P_{z}\right) $. The vertical line at $%
T=T_{0}\in \left( T_{t},T_{z}\right) $ indicates a zeroth-order phase
transition between SBHs and IBHs which is characterized by a discontinuity
in the Gibbs energy. In this area of pressures and temperatures, black hole
undergoes a first-order SBH-LBH phase transition as well. This behavior is
known as the RPT. Note that IBHs are macroscopically similar to large ones,
and thus, black holes enjoy the large-small-large phase transition in this
region of pressures and temperatures. Finally, we have just LBHs for $P<P_{t}
$ and $T<T_{t}$.

Figure \ref{PTfig} describes the coexistence lines of SBHs+LBHs (the blue
line) and IBHs+SBHs (the green line) in different scales. The blue line is
located between the critical point ($T_{c}$,$P_{c}$) and the triple point ($%
T_{t}$,$P_{t}$) between SBHs, IBHs, and LBHs. Similarly, the green line is
bounded between this triple point and point ($T_{z}$,$P_{z}$). The black
holes enjoy a first (zeroth)-order phase transition from SBHs to LBHs (IBHs
to SBHs) whenever they cross the blue (green) line from left to right or top
to bottom. Therefore, we observe the RPT behavior of the ARN black holes for
a narrow range of temperatures $T\in \left( T_{t},T_{z}\right) $\ and
pressures $P\in \left( P_{t},P_{z}\right) $.

Now, it is worthwhile to do a comparison between the ARN black holes and the
RN ones to see how this small perturbation in the Maxwell field changes the
thermodynamical behavior of the RN black holes significantly. Indeed,
observing the RPT for this kind of black hole is very interesting since such
a behavior cannot be seen for a large class of black holes even with more
complicated generalizations in the matter field and/or gravitational sector
of the field equation.

\begin{figure}[tbp]
$%
\begin{array}{ccc}
\epsfxsize=7.5cm \epsffile{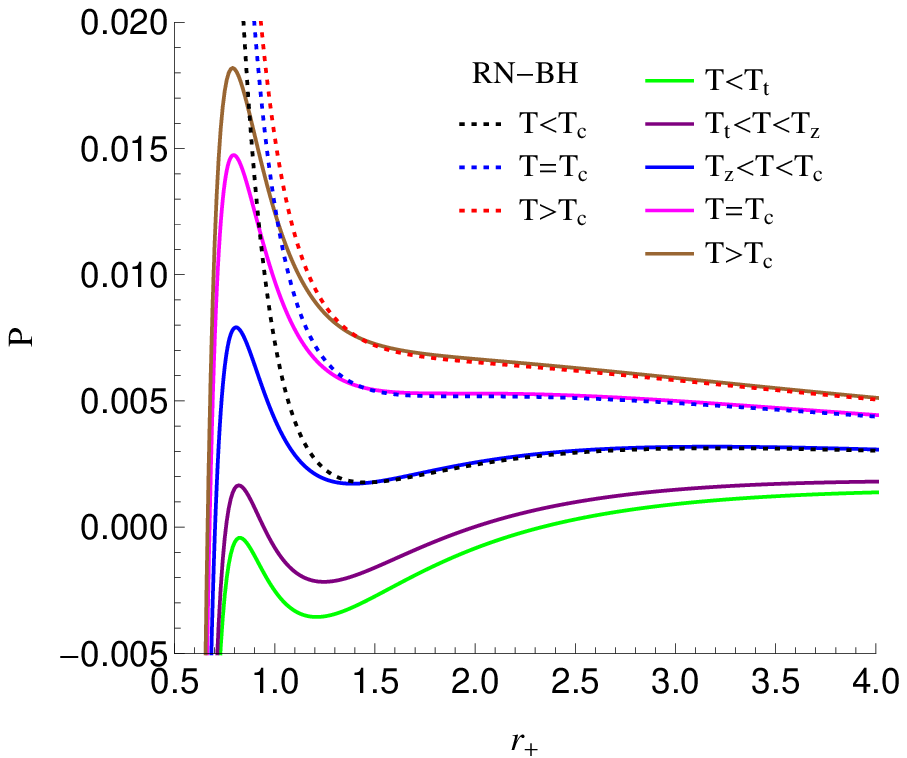} & \epsfxsize=7.5cm \epsffile{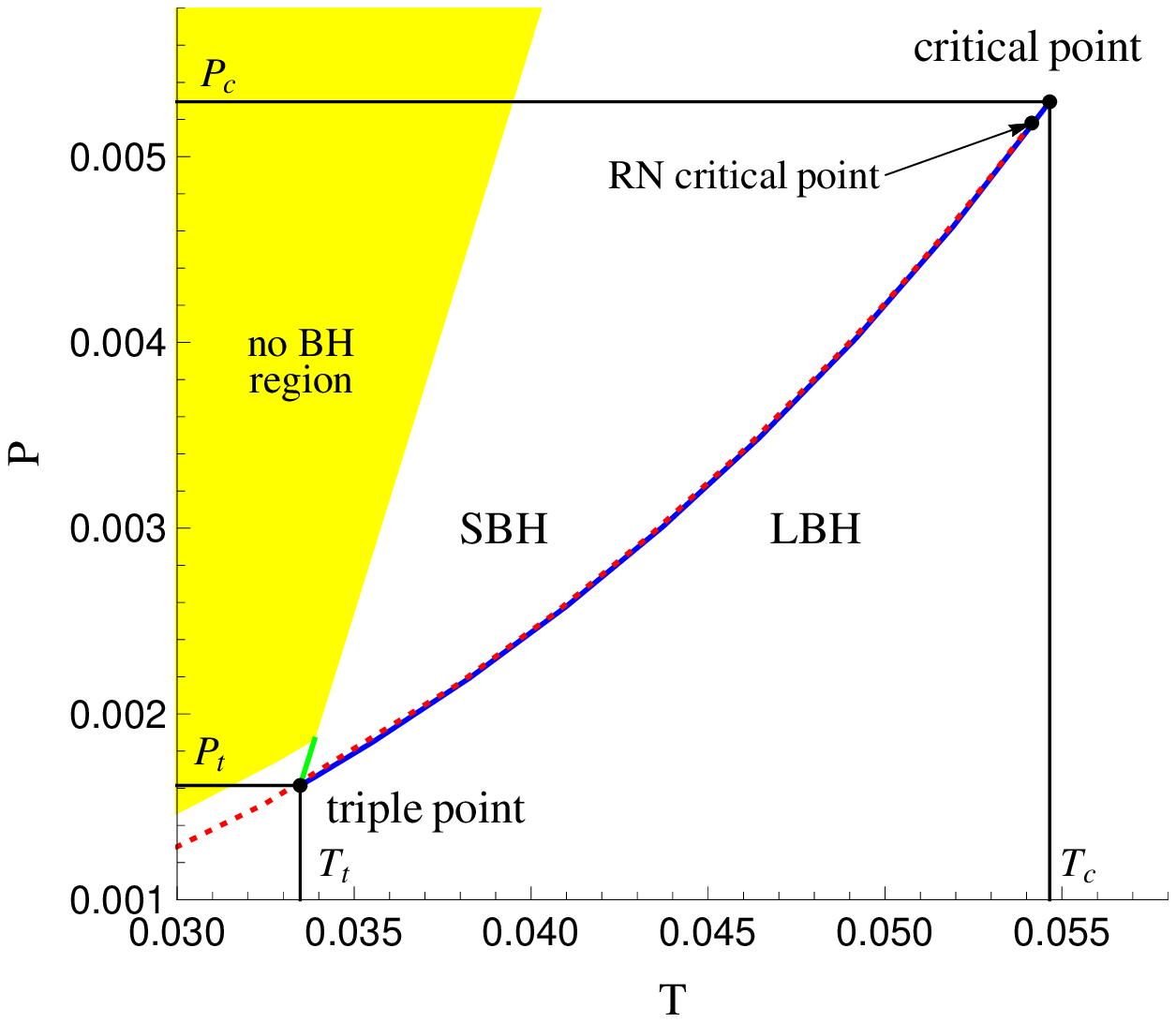}
&
\end{array}
$%
\caption{$P-r_{+}$ and $P-T$ diagrams including the RN black hole. The
dotted lines in both figures show the behavior of the RN black hole. The
nonlinear parameter highly affects the SBHs and this modification term
creates a new black hole region as IBHs and increases the critical
temperature and pressure.}
\label{comparison}
\end{figure}

Figure \ref{comparison} shows the differences between the ARN black holes
and the RN ones. From the left panel of this figure, we find that the
nonlinearity parameter reduces the pressure of SBHs significantly whereas
the pressure of LBHs almost remains unchanged. Besides, the high pressure
SBHs at low temperatures are not allowed to exist while this is not the case
for high temperature SBHs. These facts can be seen analytically from the
equation of state (\ref{P}) as well. For SBHs, the nonlinear term grows
significantly and reduces the pressure since it has a negative sign, and
finally leads to a negative pressure for these black holes that are not
allowed to exist. But for high temperatures, the first term dominates the
pressure and we have SBHs in this case. On the other hand, the correction
term will be very small for the LBHs and does not affect the pressure of
these black holes.

It is worthwhile to mention that the minimum accessible size for SBHs at
(and below) the critical point is about $r_{+}\sim 0.8$, and therefore, the
ratio of the correction and Maxwell terms ($correction/Maxwell$ ratio) is at
most about $\sim 0.3$. Thus, the nonlinear term is small even in the worst
case and never dominates the behavior of the system. In addition, from the
right panel of Fig. \ref{comparison}, one finds that the nonlinear term
creates a new region as IBHs and increases the critical temperature and
pressure. This behavior can also be understood from Eqs. (\ref{tc}) and (\ref%
{pc}) by considering the fact that the last term is ignorable since $%
r_{+c}^{4}>>2\alpha q^{2}$ (see the left panel of Fig. \ref{rhat}). However,
the nonlinearity parameter does not affect the SBH-LBH phase transition
point significantly (the right panel of Fig. \ref{comparison}).

\section{special case $\protect\alpha =4q^{2}/7$}

From the previous section, we observed that the colorful region related to $%
\alpha <4q^{2}/7$\ leads to the RPT that we have studied in details. Now, we
are interested to see what would happen for the other areas. Here, we are
going to investigate two special cases of thermodynamical behavior related
to the ARN black holes in extended phase space including a border between
the black\ and colorful areas of Fig. \ref{rhat} specified with $\alpha
=4q^{2}/7$, and also, the black area of this figure determined by $\alpha
>4q^{2}/7$. For $\alpha =4q^{2}/7$, Eqs. (\ref{rc})\ and (\ref{rbar}) give
the same results as follows%
\begin{equation}
r_{+c}=\hat{r}_{+c}=2q,  \label{rPrH}
\end{equation}%
which is a critical point since the response function (\ref{Cp}) diverges at
this point. In this case, the critical temperature (\ref{tc}) and critical
pressure (\ref{pc}) reduce to%
\begin{equation}
T_{c}=\frac{1}{7\pi q};\ \ \ \ \ P_{c}=\frac{3}{256\pi q^{2}}.  \label{tcpc}
\end{equation}

\begin{figure}[tbp]
$%
\begin{array}{ccc}
\epsfxsize=7.5cm \epsffile{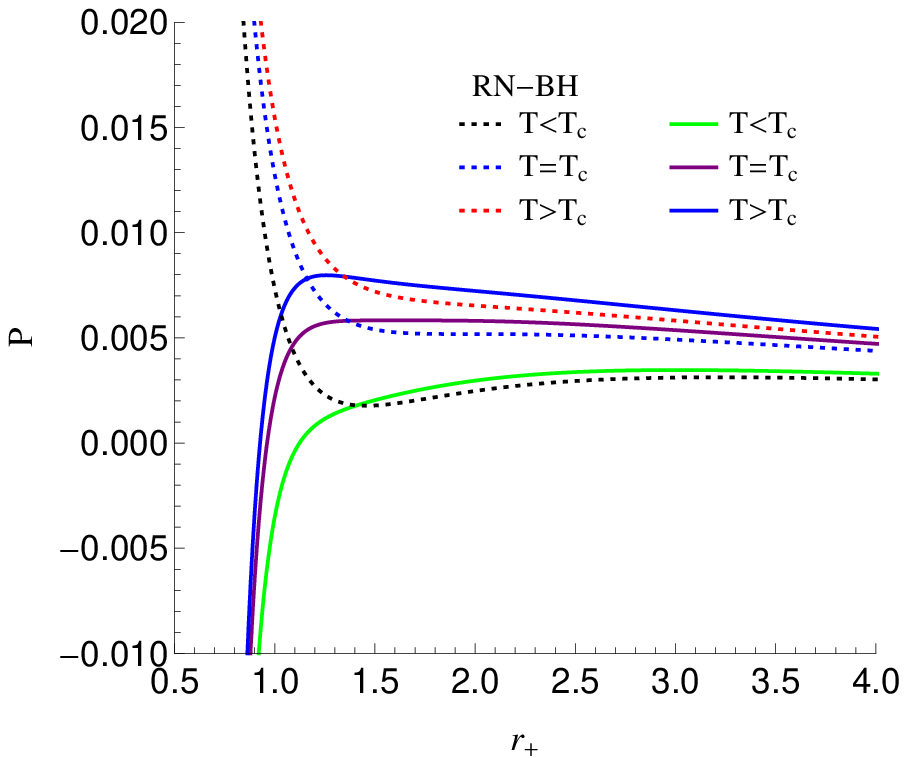} & \epsfxsize=7.5cm \epsffile{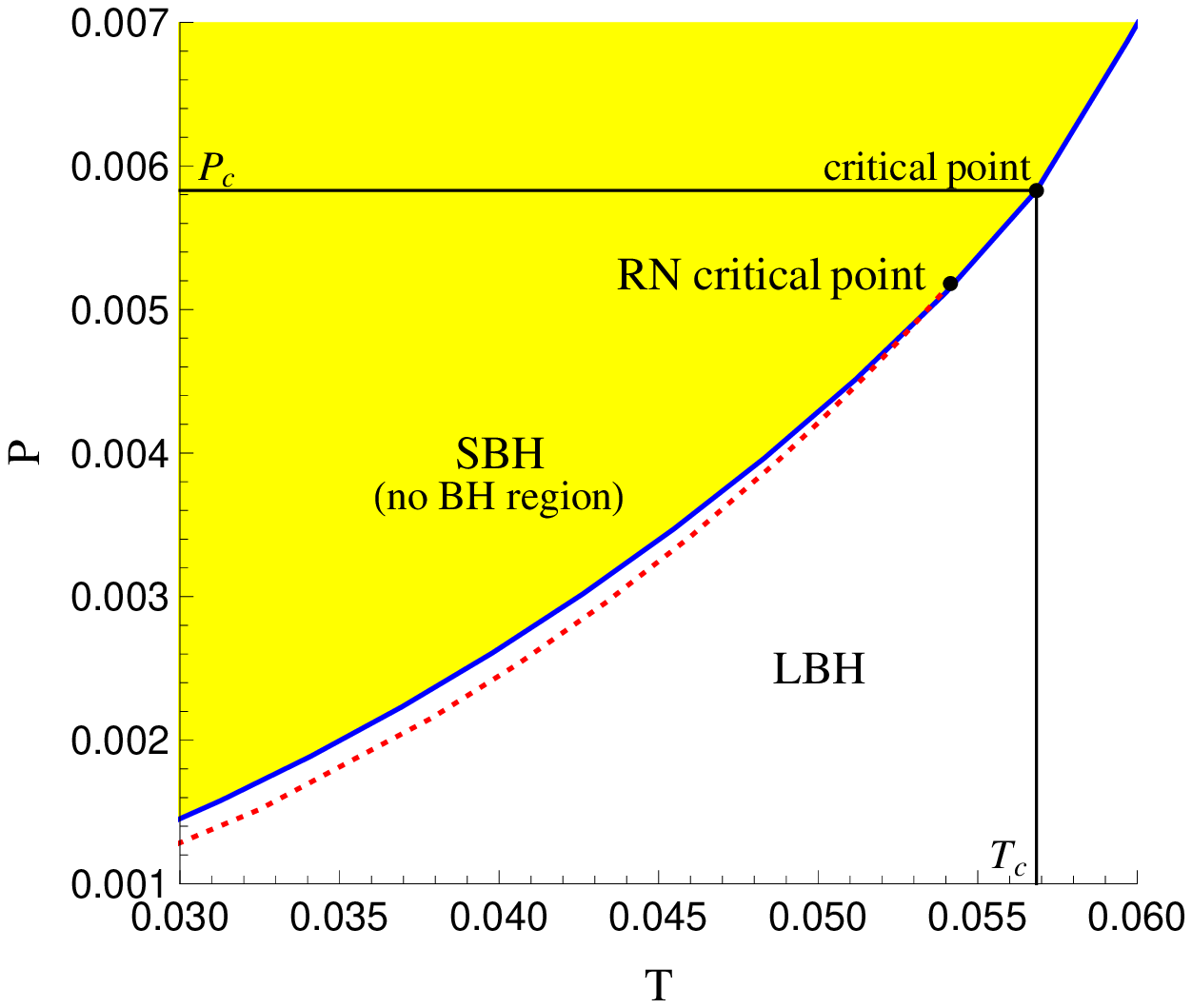}
&
\end{array}
$%
\caption{$P-r_{+}$ and $P-T$ diagrams for the special case $\protect\alpha %
=4q^{2}/7$ including the RN black hole. The dotted lines in both figures
show the behavior of the RN black hole. The nonlinear parameter converts the
stable small RN black holes to unconditionally unstable SBHs. The nonlinear
term extends the LBH region and increases the pressure of these black holes.
}
\label{SC}
\end{figure}

Here, we fix $q$ as $q=0.8$\ to plot Fig. \ref{SC}. Interestingly, this
figure indicates that the small correction highly affects the
thermodynamical behavior of the ARN black holes in the case of $\alpha
=4q^{2}/7$\ as well. In this case, the correction term is very small, hence
ignorable for all accessible black holes' event horizon radius $r_{+}$. From
the left panel, we find that the nonlinearity parameter converts the stable
small RN black holes to unconditionally unstable SBHs and slightly affects
the LBHs. The right panel shows the effect of $\alpha $ on LBHs so that the
region of these black holes is extended and the pressure is increased. In
the case of the RN black holes (and also, the critical phenomena of other
black hole solutions) the SBHs and LBHs are thermodynamically
indistinguishable above the critical point since the coexistence curve
always terminates at the critical point whereas for our black hole case
study, there is always a border between (unconditionally unstable) SBHs and
large ones (blue line of right panel of Fig. \ref{SC}). This distinguishable
property is a novel feature observed in this special type of black holes and
is due to the fact that there is no SBH-LBH phase transition in this special
case and SBHs are always unstable. Indeed, another interesting and new
behavior is that there is no SBH-LBH phase transition at (and below) the
critical point.

It is worthwhile to mention that the thermodynamic behavior of the black
area of Fig. \ref{rhat} determined by $\alpha >4q^{2}/7$, is very similar to
$\alpha =4q^{2}/7$\ case, but $r_{+c}$\ and $\hat{r}_{+c}$\ are imaginary,
and thus, the critical point specified by (\ref{rPrH}) and (\ref{tcpc})\ is
absent. Thus, in this case, the SBHs and LBHs are always distinguishable
while the SBHs are unconditionally unstable.

\section{Conclusions \label{Conclusions}}

In this paper, we have considered the cosmological constant as
thermodynamical pressure and studied the thermodynamics of $4$-dimensional
ARN black holes in the canonical ensemble of extended phase space and
deviations from those for the standard RN black holes were investigated. We
interestingly found that by considering a small correction in the Maxwell
field, the thermodynamical behavior of the RN black holes changes
significantly and a novel critical phenomenon can be observed. Based on the
values of the nonlinearity parameter, the phase space classified into three
regions, and thus, three kinds of behaviors have been found which one of
them was the RPT and the other one was a novel behavior in the extended
phase space thermodynamics.

Specially, we have seen that in addition to the standard vdW-like phase
transition of the black hole case study \cite{vdW} and the RN black holes
\cite{KubiznakMann}, they can enjoy the RPT by considering this small
correction in the Maxwell field. It was shown that this behavior happens for
a narrow range of temperatures\ and pressures. In this range of RPT, black
holes undergo a zeroth-order IBH-SBH phase transition and first-order
SBH-LBH phase transition, and this behavior could be seen for special values
of the nonlinearity parameter $\alpha <4q^{2}/7$. In comparison with the RN
black holes, the nonlinearity parameter highly affected the SBHs and
converted them to unstable ones. This modification term created a new black
hole region as IBHs and increased the critical temperature and pressure as
well.

Moreover, it was shown that for special values of the nonlinearity parameter
as $\alpha \geq 4q^{2}/7$, the correction term highly affects the
thermodynamical behavior of the solutions as well. Specially, in the case of
$\alpha =4q^{2}/7$, we observed a novel critical point such that below and
at this point, the black holes had no phase transition, and above this
critical point, SBHs and LBHs were thermodynamically distinguishable.
Besides, the stable small RN black holes converted to unconditionally
unstable SBHs. This nonlinear term extended the area of LBHs and increased
the pressure of these black holes.

As the final remark, since introducing a small correction in the Maxwell
field, interestingly, had significant effects on the thermodynamical
structure of the RN black holes, it would be nice to consider dynamical
perturbations in the background geometry of the ARN black holes and
investigate the effects of the nonlinearity parameter on the dynamical
stability and quasinormal modes, and then compare them with those of the RN
solutions.

\begin{acknowledgements}
We wish to thank Shiraz University Research Council.
\end{acknowledgements}

\end{document}